\documentstyle[aps]{revtex}
\begin{document}
\draft
\title{Relativistic static fluid spheres with a linear equation of state}
\author{B.V.Ivanov}
\address{Institute for Nuclear Research and Nuclear Energy,\\
Tzarigradsko Shausse 72, Sofia 1784, Bulgaria}
\maketitle

\begin{abstract}
It is shown that almost all known solutions of the kind mentioned in the
title are easily derived in a unified manner when a simple ansatz is imposed
on the metric. The Whittaker solution is an exception, replaced by a new
solution with the same equation of state.
\end{abstract}

\pacs{04.20.Jb}

\section{Introduction}

Einstein's equations for static and spherically symmetric perfect fluids
have been studied by many authors [1-3]. The unknown functions are more than
the equations, therefore, either an equation of state should be prescribed
or some simplifying ansatz assumed. Schwarzschild found the first solution
[4] in 1916. It was a degenerate case ($n=0$) of the linear equation of
state 
\begin{equation}
\rho =np+\rho _0  \label{1}
\end{equation}
where $\rho $ and $p$ are the density and the pressure of the fluid and $n$, 
$\rho _0$ are constant parameters. A year later, two other solutions
satisfying Eq. (1) were found, describing static cosmological models [5,6].
Tolman [7] used eight simplifying ansatze for the metric components to find
new solutions. Some limiting cases of these satisfy Eq. (1) too, leading to
a simple, but singular solution for arbitrary $n$ and $\rho _0=0$ [3]. Klein
rediscovered this solution first for the case of incoherent radiation $n=3$
[8] and then for general $n$ [9] by studying systematically the equation of
state (1) with vanishing $\rho _0$ (the so-called $\gamma $-law). This
Klein-Tolman (KT) solution was rediscovered later three times more [10-12].
Meanwhile, two new solutions with $\rho _0\neq 0$ appeared in 1968,
developing further the idea of Schwarzschild's incompressible sphere.
Whittaker argued that the gravitational mass density $\rho +3p$ should be
constant [13], while Buchdahl and Land demanded that the speed of sound
should be the largest possible, namely, equal to the speed of light [14].

The main stream of papers, however, utilized simplifying assumptions, which
often lead to complicated equations of state, if any. Already Tolman used
the ansatz 
\begin{equation}
e^{-\lambda }=a+br^2  \label{2}
\end{equation}
where $a$ and $b$ are constants, $0\leq a\leq 1$, and $e^\lambda $ is the $%
g_{rr}$ component of the metric. A thorough study of this ansatz was
performed by Kuchowicz [15,16] who found seven different cases in total,
many of them having rather involved expressions for the pressure and the
density. He didn't study whether his solutions satisfy Eq. (1).

What happens when we impose on the field equations both Eqs. (1) and (2)? An
overdetermined system of differential equations is obtained which may be
devoid of any non-trivial solutions. The real answer, derived in this paper,
is somewhat surprising. All known explicit solutions with a linear equation
of state satisfy also Eq. (2), except for the Whittaker solution. We find an
additional solution with $n=-3$ instead, which appears to be new.

It is much easier to impose both requirements directly on the field
equations, than to study case by case the results of Ref. [16]. In this way
we derive in a unified and simple manner almost all known solutions with
linear equation of state, including a new hybrid between the Einstein static
universe (ESU) and the KT solution. This programme is executed in the
following section. The last section contains some discussion.

\section{Field equations and their solutions}

The metric of a static spherically symmetric spacetime reads 
\begin{equation}
ds^2=e^\nu dt^2-e^\lambda dr^2-r^2d\Omega ^2  \label{3}
\end{equation}
where $d\Omega ^2$ is the metric on the two-sphere and $\nu $, $\lambda $
depend on $r$. The system of equations is 
\begin{equation}
\rho =\frac{\lambda ^{\prime }}re^{-\lambda }+\frac 1{r^2}\left(
1-e^{-\lambda }\right)  \label{4}
\end{equation}
\begin{equation}
p=\frac{\nu ^{\prime }}re^{-\lambda }-\frac 1{r^2}\left( 1-e^{-\lambda
}\right)  \label{5}
\end{equation}
\begin{equation}
p^{\prime }+\frac 12\left( \rho +p\right) \nu ^{\prime }=0  \label{6}
\end{equation}
where the prime means a $r$-derivative and one of the field equations has
been replaced by the contracted Bianchi identity (6). Let us impose Eq. (2).
Then Eq. (4) gives a simple expression for the density 
\begin{equation}
\rho =-3b+\frac{1-a}{r^2}  \label{7}
\end{equation}
which is regular at the centre if $a=1$. When Eq. (1) holds, Eq. (6) may be
integrated. Two different cases arise, $n=-1$ and $n\neq -1$.

Let us discuss first the case $n=-1$. Then Eq. (6) yields 
\begin{equation}
p=-\frac 12\rho _0\nu +h  \label{8}
\end{equation}
with $h$ being an integration constant. Plugging Eq. (8) into Eq. (1) and
the resulting expression for $\rho $ into Eq. (7) gives 
\begin{equation}
\frac 12\rho _0\nu =h-\rho _0-3b+\frac{1-a}{r^2}  \label{9}
\end{equation}
There are two possibilities. If $\rho _0=0$ then from Eq. (8) $p=h$, from
Eq. (1) $\rho =-p$ and from Eq. (9) $a=1$, $b=h/3$. Therefore 
\begin{equation}
e^{-\lambda }=1+\frac h3r^2  \label{10}
\end{equation}
while Eq. (5) gives $\nu $ after a simple integration: $\nu =-\lambda $.
This is the de Sitter solution.

If $\rho _0\neq 0$ then the insertion of Eqs. (2,8,9) into Eq. (5) yields
the relation 
\begin{equation}
\rho _0\left( \rho _0+2b\right) r^4=-4\left( 1-a\right) \left( a+br^2\right)
\label{11}
\end{equation}
Obviously, $a=1$ and $b=-\rho _0/2$. From Eq. (7) $\rho =3\rho _0/2$ and
from Eq. (1) $p=-\rho _0/2$. Eq. (8) gives $\nu =1+2h/\rho _0=const$, while
Eq. (2) becomes 
\begin{equation}
e^{-\lambda }=1-\frac{\rho _0}2r^2  \label{12}
\end{equation}
The pressure and the density satisfy, in fact, the equation of state $\rho
+3p=0$. Hence, we obtain ESU, although initially $\rho =-p+\rho _0$.

Let us continue with the generic case $n\neq -1$. Now the integration of Eq.
(6) brings a different answer 
\begin{equation}
p=p_1e^{-\frac{n+1}2\nu }-p_0  \label{13}
\end{equation}
where $p_1$ is a constant of integration, while $p_0=\rho _0/\left(
n+1\right) $. Proceeding like we have done with Eq. (8) the analog of Eq.
(9) is obtained 
\begin{equation}
np_1e^{-\frac{n+1}2\nu }-np_0+\rho _0=-3b+\frac{1-a}{r^2}  \label{14}
\end{equation}
Again, there are two possibilities. If $n=0$ we have $a=1$, $b=-\rho _0/3$
and from Eq. (1) $\rho =\rho _0=const$. The metric component $\nu $ is
obtained not from Eq. (14) but from Eq. (5), the last equation to be
satisfied. For a general $n$ it becomes a linear equation for $y=e^{\frac{n+1%
}2\nu }:$%
\begin{equation}
\frac 2{n+1}r\left( a+br^2\right) y^{\prime }=\left[ 1-a-\left( p_0+b\right)
r^2\right] y+p_1r^2  \label{15}
\end{equation}
In our case it reduces to 
\begin{equation}
2\left( 1-\frac{\rho _0}3r^2\right) y^{\prime }=-\frac 23r\rho _0y+p_1r
\label{16}
\end{equation}
and is easily integrated 
\begin{equation}
e^\nu =\left( \frac{3p_1}{2\rho _0}+Ce^{-\lambda /2}\right) ^2  \label{17}
\end{equation}
\begin{equation}
e^{-\lambda }=1-\frac{\rho _0}3r^2  \label{18}
\end{equation}
which, after some change of notation, coincides with solution III of Tolman.
This is the Schwarzschild interior solution.

If $n\neq 0$ both Eqs. (14) and (15) determine $\nu $ and we must study
their compatibility. Eq. (14) may be written as 
\begin{equation}
\frac 1y=A+\frac B{r^2}  \label{19}
\end{equation}
\begin{equation}
A=-\frac 1{n\left( n+1\right) p_1}\left[ \rho _0+3\left( n+1\right) b\right]
\label{20}
\end{equation}
\begin{equation}
B=\frac{1-a}{np_1}  \label{21}
\end{equation}
When these relations are put into Eq. (15), a polynomial of second degree in 
$r^2$ arises, which must vanish identically. Three compatibility conditions
result 
\begin{equation}
\left( 1-a\right) \left[ \frac{4n+\left( n+1\right) ^2}{\left( n+1\right) ^2}%
a-1\right] =0  \label{22}
\end{equation}
\begin{equation}
A\left[ \rho _0+\left( n+3\right) b\right] =0  \label{23}
\end{equation}
\begin{equation}
B\left[ \left( n+1\right) \rho _0+4nb+\left( n+1\right) \left( n+3\right)
b\right] -\left( n+1\right) ^2\left( 1-a\right) A=0  \label{24}
\end{equation}
Eq. (22) has two solutions for $a$. Let $a=1$. Then $B=0$, $\nu $ is a
constant and $\rho =-3b$. Eq. (24) is also satisfied, while Eq. (23)
provides two subcases. If $A=0$ then from Eq. (19) $y=\infty $ which is
unacceptable. If $\rho _0=-\left( n+3\right) b$ then Eq. (20) shows that $%
A\neq 0$ when $b\neq 0$. Eq. (1) gives $p=b$, hence the equation of state is 
$\rho +3p=0$ for any $n$. We again obtain the ESU with constant $\nu $ and 
\begin{equation}
e^{-\lambda }=1+br^2  \label{25}
\end{equation}
This is the same as Eq. (12), since both can be written as $e^{-\lambda
}=1+pr^2$.

The last case which remains is 
\begin{equation}
a=\frac{\left( n+1\right) ^2}{4n+\left( n+1\right) ^2}  \label{26}
\end{equation}
where $a\neq 1$ since $n\neq 0$. Eq. (23), like before, offers two choices.

a) $A=0$ ($\rho _0=-3\left( n+1\right) b$). Then Eq. (24) simplifies to $%
\left( n-1\right) b=0$. There are two subcases:

a1) $n=1$, $b\neq 0$. Then $\rho _0=-6b$, Eq. (26) gives $a=1/2$, Eq. (21)
gives $B=1/2p_1$ and Eqs. (2,19,7,1) give respectively 
\begin{equation}
e^{-\lambda }=\frac 12\left( 1-\frac{\rho _0}3r^2\right)   \label{27}
\end{equation}
\begin{equation}
e^\nu =2p_1r^2  \label{28}
\end{equation}
\begin{equation}
\rho =\frac 12\left( \rho _0+\frac 1{r^2}\right)   \label{29}
\end{equation}
\begin{equation}
\rho =p+\rho _0  \label{30}
\end{equation}
This is precisely the solution of Buchdahl and Land [14] after $e^\nu $ is
rescaled.

a2) $b=0$. Then $\rho _0=0$ and we easily find 
\begin{equation}
e^{-\lambda }=\frac{\left( n+1\right) ^2}{4n+\left( n+1\right) ^2}
\label{31}
\end{equation}
\begin{equation}
e^{-\frac{n+1}2\nu }=\frac 4{\left[ 4n+\left( n+1\right) ^2\right] p_1r^2}
\label{32}
\end{equation}
\begin{equation}
\rho =np=\frac{4n}{\left[ 4n+\left( n+1\right) ^2\right] r^2}  \label{33}
\end{equation}
This is the KT solution in curvature coordinates [3,8-12]. Eq. (31) shows
that the solution exists when $4n+\left( n+1\right) ^2>0$. This condition is
fulfilled for $n$ outside the interval $\left( -5.83=-3-2\sqrt{2},-3+2\sqrt{2%
}=-0.17\right) $. When $n<-5.83$ or $-0.17<n<0$ the solution has $p$ and $%
\rho $ of different signs. The physically realistic range is $1\leq n\leq
\infty $, which includes the important cases of stiff fluid ($n=1$) and
incoherent radiation ($n=3$). The KT solution is singular at $r=0$ but it
can serve as an interior solution for $r>r_0>0$ in multi-layered perfect
fluid models. It also represents a focal point for a regular solution, which
has been studied numerically [8,17,18, 19].

b) $\rho _0=-\left( n+3\right) b$. Since $a\neq 1$, Eq. (24) reduces to $%
\left( n+3\right) b=0$. The two subcases are

b1) $b=0$. Then $A=0$ and we go back to case a).

b2) $n=-3$. Then $\rho _0=0$, $a=-1/2$, $\rho +3p=0$ and 
\begin{equation}
e^{-\lambda }=-\frac 12+br^2  \label{34}
\end{equation}
\begin{equation}
p_1e^\nu =b-\frac 1{2r^2}  \label{35}
\end{equation}
\begin{equation}
\rho =-3b+\frac 3{2r^2}  \label{36}
\end{equation}
The constant $p_1$ may be set to one by a time rescaling. This appears to be
a new solution - a hybrid between ESU with its constant pressure and density
and the KT solution. Its pressure, density and metric component $g_{tt}$ are
linear combinations of those two solutions, although Eqs. (4-6) form a
highly non-linear system. When $b=0$ we obtain the KT solution for $n=-3$
which, in fact, does not exist. When $b\neq 0$, $e^\lambda $ and $e^\nu $
are negative for small $r$. However, if we choose $r_0^2=1/2b$ and $b>0$,
then the solution exists for $r>r_0$, has positive pressure, negative
density and represents a regular companion of ESU.

Whittaker has solved the general case $n=-3$ when $\rho _0$ may be different
from zero and the ansatz (2) is not imposed [13]. However, the above
solution has been missed. The reason is the following. Eqs. (1,4,5) lead to
the relation 
\begin{equation}
\left( e^{\lambda +\nu }\right) ^{\prime }=kre^{2\left( \lambda +\nu \right)
}  \label{37}
\end{equation}
where $k$ is a constant. The integration of this equation yields 
\begin{equation}
e^{\lambda +\nu }=\frac 2{c-kr^2}  \label{38}
\end{equation}
$c$ being a constant. Whittaker assumes that $k=c\alpha $, where $\alpha $
is yet another constant. This choice excludes the above solution which has $%
c=0$, $k=-2p_1\neq 0$ and in principle may be regained as a limiting case.
However, when one sets $\rho _0=0$ in the results of Ref. [13] only ESU is
obtained. It should be noticed that the singular term in Eq. (35) is
different from the universal Schwarcschild term $m/r$ which was correctly
discarded from the interior solution.

\section{Discussion}

In this paper we have studied the gravitation of static spherically
symmetric perfect fluid solutions with a linear equation of state. We
imposed the additional requirement (2), making the system of equations
overdetermined. Amazingly, it gives six non-trivial solutions: the
Schwarzschild interior solution for an incompressible sphere, the de Sitter
cosmological solution, the Einstein static universe, the stiff fluid
solution of Buchdahl and Land, the Klein-Tolman singular solution and a new
solution with $\rho +3p=0$. Among the known solutions with linear equation
of state only the Whittaker solution does not satisfy Eq. (2). The reason is
that it satisfies Eq. (38) and if Eq (2) is required too, $\nu $ has the
only option to remain a constant, which is too restrictive. These solutions
have been found originally starting from a variety of physical
considerations and following quite different mathematical approaches. Here
they are derived in a unified way as a result of straightforward
calculations and a new solution is added. Roughly speaking, it is a linear
combination between ESU and the KT solution, which is unusual for a
non-linear system of differential equations. It closes the list of solutions
satisfying both Eqs. (1) and (2). The equation of state $\rho +3p=const$ is
unrealistic because the speed of sound is negative. However, models which
satisfy it have interesting properties, are often integrable and occupy
important places in different classification schemes. This concerns one
whole class of algebraically special perfect fluids of Petrov type II or D
[1] (Eq.(29.20)), a rigidly rotating solution of type D [20], the well-known
Wahlquist solution [21-22] which has the Whittaker solution as a static
limit and the Kerr metric as a vacuum subcase, and the recently found
Wahlquist-Newman solution [23] which reduces to a number of well-known
solutions in appropriate limits.  An interesting phenomenon is the effective
change of the equation of state by which ESU is obtained first from $\rho
=-p+\rho _0$ and then from $\rho =np+\rho _0$. When $\rho _0=0$ a
one-parameter sequence of KT solutions exists which serve as focal points
for regular numerical solutions. When $\rho _0\neq 0$ only two discrete
cases satisfy Eq. (2) for $n=0$ and $n=1$, the third known one ($n=-3$)
drops out of this scheme. Presumably when one replaces the ansatz (2) by a
different one, other integrable cases will appear, although the system is
overdetermined.

\end{document}